\documentclass[12pt,preprint]{aastex}
\slugcomment{To appear in ApJ Letters: 10 November 2006}

\def\la{\lower.5ex\hbox{$\; \buildrel < \over \sim \;$}}
\def\ga{\lower.5ex\hbox{$\; \buildrel > \over \sim \;$}}

\begin{document}

\title{Infrared dust emission in the outer disk of M51}

\author{Michele D. Thornley}
\affil{Department of Physics \& Astronomy, Bucknell University, Lewisburg, PA 17837}
\email{mthornle@bucknell.edu}

\and

\author{Jonathan Braine and Erwan Gardan} 
\affil{Observatoire de Bordeaux, UMR 5804, CNRS/INSU, B.P. 89, F-33270 Floirac,
France}
\email{braine@obs.u-bordeaux1.fr, gardan@obs.u-bordeaux1.fr}

\begin{abstract}

We examine faint infrared emission features detected in {\it Spitzer
Space Telescope} images of M51, which are associated with atomic
hydrogen in the outer disk and tidal tail at R$\ga$R$_{25}$
(4.9\arcmin, $\sim$14 kpc at d=9.6 Mpc).  The infrared colors of these
features are consistent with the colors of dust associated with star
formation in the bright disk.  However, the star formation efficiency
(as a ratio of star formation rate to neutral gas mass) implied in the
outer disk is lower than that in the bright disk of M51 by an order of
magnitude, assuming a similar relationship between infrared emission
and star formation rate in the inner and outer disks.

\end{abstract}


\section{Introduction}

Relatively little is known about the conditions in the interstellar
medium (ISM) in the outer disks of spiral galaxies, yet an accounting
of the ISM in these regions is integral to a full understanding of the
galactic star formation process.  Studies of the ISM at large radii
can measure the available reservoir of neutral gas and dust in
galaxies and assess the efficiency of ongoing star formation.
Sensitive atomic hydrogen (HI) observations often reveal extended
distributions of the ISM that may support the formation of additional
stars or tidal dwarf galaxies (see classic reviews by van der Kruit \&
Allen 1978; Haynes, Giovanelli, \& Chincarini 1984).  However,
measures of star formation associated with this extended neutral
component are rare, thus hampering efforts to fully characterize the
local star formation process.

Infrared imaging by the {\it Spitzer Space Telescope} (Werner et
al. 2004) with the IRAC (Fazio et al. 2004) and MIPS (Rieke et
al. 2004) cameras provides a new opportunity for mapping out dust in
emission in nearby galaxies.  Of particular interest is dust emission
thought to arise from stochastically-excited aromatic molecules, such
as Polycyclic Aromatic Hydrocarbons (PAHs, see, e.g., Leger \& Puget
1984; Puget \& Leger 1989; Boulanger et al.  1998; Helou et al. 2000;
Lu et al. 2003), at $\sim$8\micron\ and thermal dust emission at
$\sim$24\micron.  Studies with ISO and Spitzer suggest that 8 and
24\micron\ emission can be used to measure star formation rates on
kiloparsec to global scales (e.g., Roussel et al. 2001; Dale et
al. 2005; Calzetti et al. 2005).  In this Letter, we examine infrared
dust emission in the outer disk of M51, and compare the star forming
environment they represent with that of the bright disk of M51.

\section{Observations and Data Analysis}

\begin{figure} 
\includegraphics[width=5in]{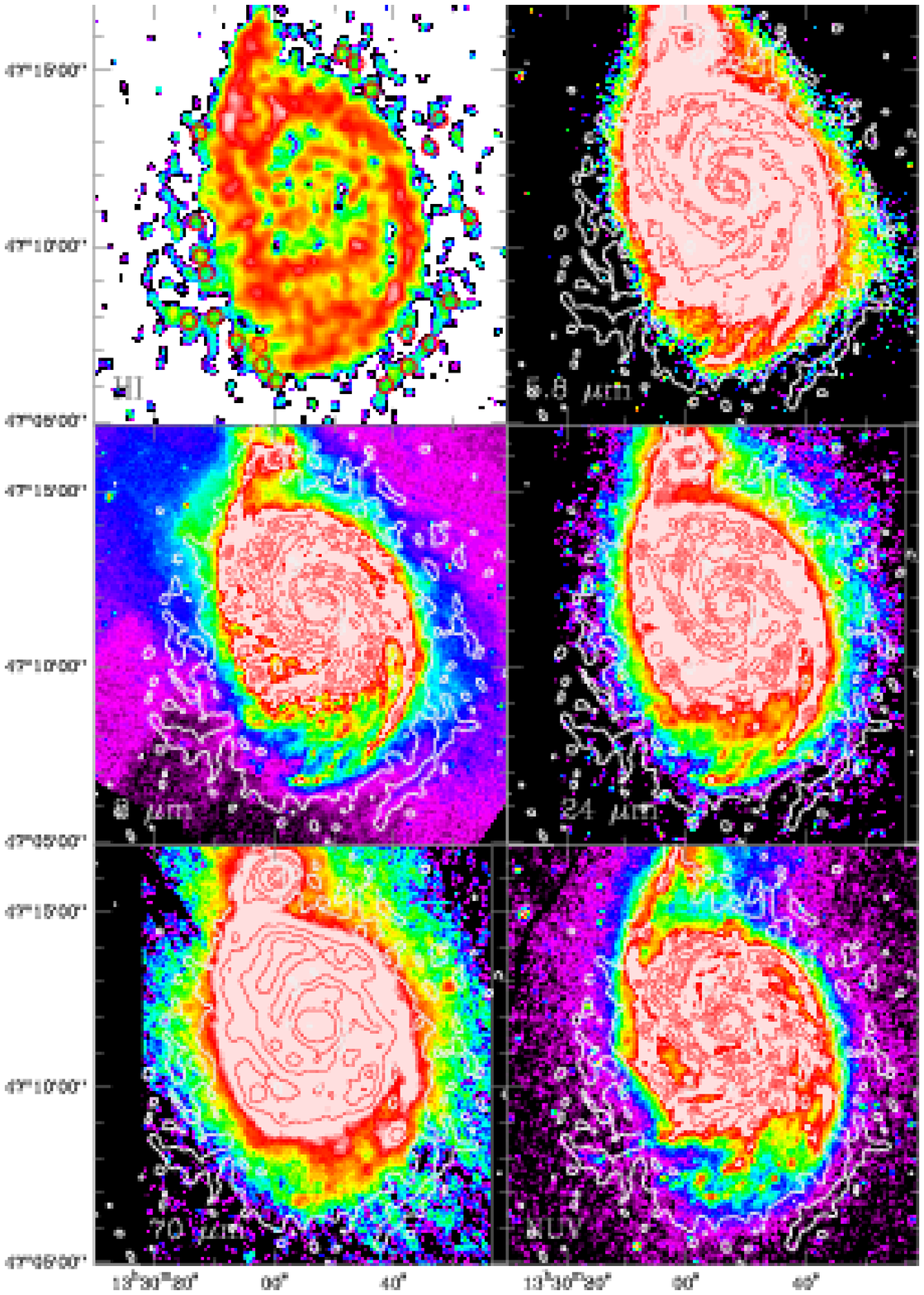}
\caption{Emission from M51 at the wavelengths listed in the lower left
  corner of each panel. Red circles with a diameter equal to the
  21\arcsec\ resolution are overlaid on the HI image, for each of the
  positions listed in Table \ref{tab1}.  The lowest flux level shown
  in the HI map (corresponding to 1.6 x 10$^{20}$ cm$^{-2}$) is
  indicated by a single white contour in the other five panels.  The
  bright inner disk in the IR and NUV images has been saturated here
  in order to show faint, outer features more clearly; red contours
  show the general structure in the inner disk at each wavelength.
  The range of fluxes displayed logarithmically by the color table in
  each image is as follows: for 5.8$\mu$m: 0.003-0.2 MJy sr$^{-1}$;
  for 8.0$\mu$m: 0.3-2.5 MJy sr$^{-1}$; for 24$\mu$m: 0.04-1.0 MJy
  sr$^{-1}$; for 70$\mu$m: 0.3-10 MJy sr$^{-1}$; for NUV:
  1.2x10$^{22}$-1.9x10$^{23}$ erg s$^{-1}$ Hz$^{-1}$.
\label{fig1}}
\end{figure}

Figure \ref{fig1} shows M51 in HI, infrared (IR), and near-ultraviolet
(NUV) emission. The Spitzer IR images (5.8, 8.0$\mu$m: IRAC; 24,
70$\mu$m: MIPS) were acquired from the Second Enhanced Data Release
(2005 May 6) of the Spitzer Infrared Nearby Galaxies Survey (SINGS,
Kennicutt et al. 2003).  The data reduction pipeline is described in
Regan et al. (2004).  Calibration uncertainties are estimated to be
10\% for the 5.8, 8.0, and 24$\mu$m bands and 20\% for the 70$\mu$m
band. No extended-source aperture corrections have been applied to the
IRAC data.  The inherent angular resolution is 2\arcsec\ for the IRAC
images, 6\arcsec\ for the 24\micron\ MIPS image, and 18\arcsec\ for
the 70\micron\ MIPS image.

The GALEX NUV (1750-2750~\AA) image was taken as part of the GALEX
Nearby Galaxies Survey (NGS; see Bianchi et al. (2005)) and acquired
from the Multimission Archive at Space Telescope. The inherent angular
resolution of the GALEX NUV image is 6\arcsec. As an additional check
on the star formation rates in the outer disk, we conducted a parallel
analysis using the H$\alpha$~ image of M51 published by Rand (1992),
which has an angular resolution of $\sim$2\arcsec.

The HI data presented here were published by Rots et al. (1990).  We
are using their 21\arcsec\ resolution dataset. To highlight the
strongest HI features in the outer disk, the integrated intensity map
shown in Figure \ref{fig1} was created using the AIPS task MOMNT, with
three-channel Hanning smoothing in velocity and five-pixel (FWHM)
Gaussian spatial smoothing. By comparing with the 34\arcsec\
resolution images shown in Rots et al.(1990), it is clear that the
HI image in Figure \ref{tab1} shows the high-column-density ridges in
the broader HI tail that extends away from the galaxy toward the southeast.


By examination of the HI and IR images, 21 positions were chosen to
represent the extended emission.  These positions, listed in Table 1,
correspond to local HI peaks which lie at intervals spaced by
$\sim$1\arcmin along the extended IR features. Sixteen positions
lie outside the m$_B$=25 mag arcsec$^{-2}$ isophote, and the remaining
five lie at radii $\ge$0.7R$_{25}$ (R$_{25}$=4.9\arcmin,LEDA \footnote{The Lyon/Meudon Extragalactic Database,
http://leda.univ-lyon1.fr/}).

HI spectra at each of the 21 positions were extracted from the
original 21\arcsec\ datacube.  The typical velocity widths
of the HI spectra are relatively small (no more than 3-4 channels for
a typical spectrum, or $\Delta V_{\rm fwhm}\sim$25~km~s$^{-1}$),
suggesting the emission comes from relatively cool, dense atomic gas.
We measured the HI integrated intensities directly from the spectra,
integrating over the observed spectral line after subtracting off a
linear baseline. The integrated intensities for all positions are
within 17\% of the integrated intensities which come from a fitted
gaussian function, and the majority are within 10\%.  The integrated
intensities were then converted to HI column densities, assuming the
standard optically-thin conversion (van de Hulst, Muller, \& Oort
1954), and reported in Table \ref{tab1}.

Some fraction of the 5.8 and 8\micron\ fluxes is due to starlight
(see, e.g., Pahre et al. 2004a,b; Helou et al. 2004).  We used the
starlight-subtraction technique of Pahre et al. (2004a), to create
images of the dust emission in M51 at 5.8 and 8.0\micron; thus, the
5.8 and 8.0\micron\ images in Figure 1 represent infrared dust
emission at these wavelengths. Starlight contributes the majority of the observed
5.8\micron\ flux at some positions, but is generally within the
background uncertainty at 8\micron.  For the 70\micron\ images, the
fluxes at each position are dominated by systematic uncertainties due
to residual striping in the image.  Due to the significantly increased
uncertainties in measured dust emission at 5.8 and 70\micron, we will
use only 8 and 24\micron\ fluxes for calculating infrared star
formation measures (\S\ref{sfrs}).

The IR, NUV, and H$\alpha$~ images were smoothed by gaussian functions
to provide 21\arcsec\ effective resolution($\sim$ 1 kpc for an assumed
distance of 9.6 Mpc), and all fluxes are consequently reported in
units of flux per 21\arcsec\ FWHM gaussian beam for consistency with
the information available in the HI image. To measure the flux at each
position, the local background was measured in 20 positions outside
the minimum HI contour shown in Figure 1, but within 2.5\arcmin\ of
the measured source position. The variance of the local background
measurements is used to represent the uncertainty in the flux at each
source position.

The background-subtracted fluxes for the smoothed 8.0\micron,
24\micron, NUV, and H$\alpha$~ images are listed in Table 1.  Fluxes
smaller than three times the measured background variance are reported
as upper limits.  In addition, we report in Table 1 the value of the
total IR luminosity, L(IR)=L(3-1100$\mu$m), as inferred from the 8 and
24$\mu$m fluxes and the empirical relation determined for M51 by
Calzetti et al. (2005)\footnote{log~L(IR)~=~log~L(24)+0.908+0.793(log
(L$_{\nu}$(8)/L$_{\nu}$(24)))}.

\section{Infrared emission in the outer disk}

The lowest HI column density shown in Figure 1 (N$_{HI}$=1.6x10$^{20}$
cm$^{-2}$) is overlaid as a single white contour on the IR and NUV
images to show the spatial relationship of the outer IR, NUV, and HI
morphologies.  Most of the IR and NUV images shown in Figure 1 have
been slightly smoothed (5.8 and 8\micron\ to 5\arcsec, 24\micron\ and
NUV to 7\arcsec) to improve the display of the faintest features.

Figure 1 shows that some, although not all, of the strongest outer HI
features are associated with locally enhanced IR emission.  Note
particularly the ridge-like feature to the southwest, which is
detected in the 5.8, 8, and 24\micron\ images, and the southeastern
extension, seen most strongly in 8 and 24\micron\ emission.  The
70\micron\ emission morphology is also consistent with the southwest
ridge, though with relatively large systematic uncertainties.
Features corresponding to the extended IR features are significantly
less prominent in the NUV image with respect to the bright disk,
though there is clearly NUV emission associated with the southeastern
IR extension.  NUV emission from the southwest ridge may be
preferentially extincted by the morphology of the ISM pulled out by
the interaction of NGC~5194 and NGC~5195, or the weak NUV emission may
simply indicate relatively little star formation at these positions.
 
\section{Infrared star formation rates}\label{sfrs}

The detection of IR dust emission associated with enhanced HI column
densities suggests that there are localized HI peaks in the outer disk
that are associated with star formation.  Here, we use a calibration
from a recent IR study of star formation tracers in the bright disk of
M51 (Calzetti et al. 2005) to self-consistently examine the star
formation process in the inner and outer disk of M51.

We have estimated the local IR star formation rate (SFR) on
$\sim$1~kpc scales at each outer disk position using the total
infrared luminosity, L(IR), as calibrated by Calzetti et al. (2005),
and the linear conversion between L(IR) and SFR of Kennicutt (1998).
Recent studies suggest that $\sim$8\micron\ IR emission is a
non-linear tracer of star formation, but the relationship between
L(IR) and SFR is nearly linear (Kewley et al. 2002, Calzetti et
al. 2005).  For this analysis, we will assume a linear conversion
between L(IR) and SFR. For comparison, H$\alpha$ fluxes at the same 21
positions were used to determine H$\alpha$ SFRs.  The H$\alpha$ fluxes
were corrected for extinction using the measured HI column densities,
assuming that the dust associated with the gas has Galactic properties
and is distributed in a foreground screen.  The conversion factor from
Kennicutt (1998) was then used to obtain an H$\alpha$ SFR.

The IR and H$\alpha$ SFRs are listed in Table 2, and plotted as a
function of HI column density in Figure 2.  There is no discernable
trend in SFR with HI column density, but the range of HI column
density (and radius) is small.  It is clear that we are detecting real
infrared flux, as seen by the correlation between 8 and 24\micron\
fluxes shown in the inset of Figure 2.  In addition, the
8\micron/24\micron\ flux ratios are consistent with values measured
within the optical disks of M51 and other nearby spiral galaxies
(Regan et al. 2004; Helou et al. 2004; Calzetti et al. 2005), and are
high enough to suggest that depletion of metals will not contribute to
a dearth of PAH-band emitters at 8\micron\ (Engelbracht et al. 2005).
This is particularly true for M51, where there is little or no radial
metallicity gradient (Bresolin, Garnett, \& Kennicutt 2004).  The
H$\alpha$ SFRs are characteristically lower, but except for position
3, they are within a factor of five of the IR SFRs at the positions
where both IR and H$\alpha$ SFRs are measured.

The L(IR) SFRs per 21\arcsec~beam in the outer disk positions are
calculated to be (2-20)x10$^{-4}$ M$_\odot$~yr$^{-1}$, and the
H$\alpha$ SFRs lie in the range (1-6)x10$^{-4}$ M$_\odot$~yr$^{-1}$.
The corresponding gas masses per beam inferred from the HI column
density (including a factor of 1.36 for helium) are
$\sim$(3-12)x10$^{6}$ M$_\odot$, with a median value of 6x10$^{6}$
M$_\odot$.

Ten positions in the bright inner disk were chosen to represent
typical inner disk SFRs. The SFRs determined from H$\alpha$ and L(IR)
from these positions are also listed in Table 2.  The L(IR) SFRs for
the inner disk positions are typically within a factor of three of the
H$\alpha$ SFRs, consistent with the measured scatter in the empirical
relationship between L(IR) and the 8/24 ratio in the bright inner disk
(Calzetti et al. 2005).  Again, the inferred H$\alpha$ SFRs are
generally lower than SFRs determined from L(IR), but both are two
orders of magnitude higher than in the outer disk: we measure
(160-770)x10$^{-4}$ M$_{\odot}$ yr$^{-1}$ using L(IR), and
(130-660)x10$^{-4}$ M$_{\odot}$ yr$^{-1}$ using H$\alpha$.  The inner
disk gas masses per beam were $\sim$(12-25)x10$^{6}$ M$_\odot$, as
determined from a combination of the HI column density and the H$_2$
column density. N(H$_2$) was measured from the
published CO integrated intensity maps published by Helfer et
al. (2003), assuming N(H$_2$)=(3x10$^{20}$cm$^{-2}$(K km
s$^{-1}$)$^{-1}$)I$_{CO}$.

Taking the ratio of the SFR and the available gas mass in the same
beam, we find the star formation efficiency (SFE) for each position
(Table 2).  For the outer disk positions of M51, the L(IR) SFEs are
0.04-0.2~Gyr$^{-1}$, and the H$\alpha$ SFEs are 0.02-0.06~Gyr$^{-1}$.
The median SFE for the 10 outer disk positions where significant
emission was detected in H$\alpha$, 8\micron, and 24\micron\ emission
is 0.04 and 0.07~Gyr$^{-1}$ from H$\alpha$ (corrected) and L(IR),
respectively.  For the inner disk positions of M51, the L(IR) SFEs are
0.8-4~Gyr$^{-1}$, and the H$\alpha$ SFEs are 0.6-3~Gyr$^{-1}$.  The
median SFE for the inner disk positions is 1.3 and 2.8~Gyr$^{-1}$ from
extinction-corrected H$\alpha$ and L(IR), respectively.

A comparison of the median SFEs for inner and outer disk positions
suggests that star formation is $\sim$30-40 times less efficient in
the outer disk than in the bright disk, presuming that 8 and
24\micron\ emission traces star formation similarly in the inner and
outer disk.  While the uncertainties in these measures are large in
considering both measures of low flux values and significant
assumptions in the local star formation calibration, these
calculations suggest that star formation is roughly an order of
magnitude less efficient in the outer disk than in the bright inner
disk.
 
\section{Robustness of infrared star formation efficiencies}\label{robust}

To assess the robustness of our conclusion of low SFE in the outer
disk of M51, we also compared the H$\alpha$ star formation rates
calculated here with rates determined from NUV emission, using the
NUV-SFR calibration of Kennicutt (1998). The calculated SFRs from NUV
measures are also listed in Table 2.  For the positions where both NUV
and H$\alpha$ flux is detected significantly, the derived SFRs
(without extinction correction) are within a factor of $\sim$2.5 of
one another.  Given the relatively small extinction corrections at these
positions, these measures are expected to be reasonably representative
of the outer disk star formation rate.  This comparison, combined with the
H$\alpha$-IR comparison above, suggests that the IR measurements
presented here do not artificially underestimate the SFR (e.g., due to
potentially lower dust temperatures and metallicities in the outer
disk).

In considering the robustness of our relative SFE measures, we must
also assess the accuracy of our neutral gas measures. Molecular gas is
likely present in the outer disk, but with a significantly lower
average column density (see, e.g., Braine \& Herpin 2004) over
$\sim$1~kpc scales, such that the contribution of molecular mass to
the total neutral gas mass should be small.  In any case, a
contribution from molecular gas in the outer disk only creates a
larger discrepancy between the SFE in the inner and outer disks of
M51.  Similarly, if the CO-to-H$_2$ conversion factor used for the
inner disk is too high, the discrepancy between inner and outer disk
SFEs would also increase.

\section{Discussion and Conclusions}

Measures of IR, UV, and H$\alpha$ emission associated with the
extended neutral gas component of M51 suggest that star formation in
the outer disk of M51 is roughly an order of magnitude less efficient
than at smaller radii.  This may be due to slower HI to H$_2$
conversion, possibly because of the lower pressures in the outer disk
ISM.  Recent CO and UV observations (Braine \& Herpin 2004; Neff et
al. 2005) have shown that both molecular gas and star formation are
present in the outer disks of spiral galaxies, and they appear linked
to the HI.  Deeper infrared studies of the outer regions of spirals
will provide a valuable opportunity to test the local effects of the
interstellar radiation field, metallicity, and gas phase on infrared
dust emission, and thus help to better define the relationship of star
formation and gas content.

\acknowledgments MDT would like to acknowledge the warm hospitality of
the Observatoire de Bordeaux, where this work was carried out.  The
authors thank S. Vogel and A. Rots for providing the HI data and
R. Rand for providing his published H$\alpha$ map.

\clearpage

\clearpage

\begin{figure} 
\includegraphics[width=3.5in]{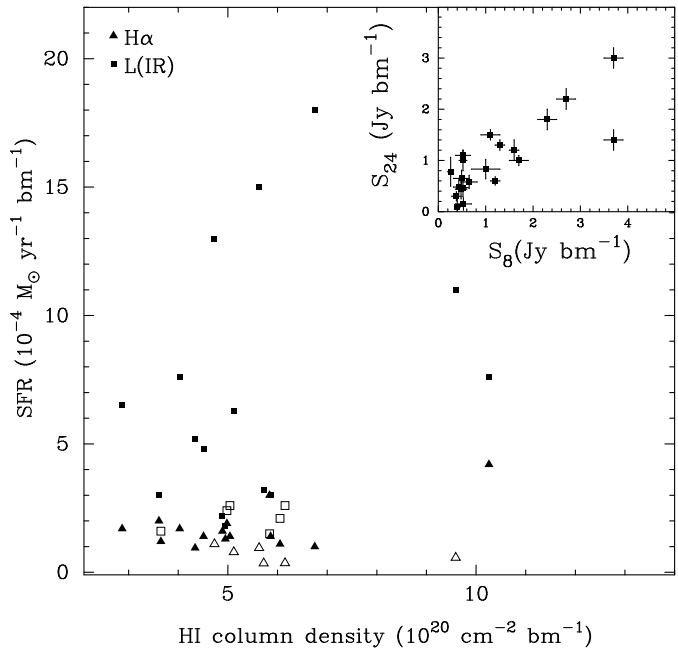}
\caption{Star formation rates, from H$\alpha$ (triangles) and
  L(IR) (squares), for the 21 positions from Table \ref{tab1}, plotted as a
  function of HI column density. Upper limits are indicated by open symbols.
  Inset: 8\micron\ versus 24\micron\ flux for the 21 outer disk
  positions. 
  \label{fig3}}
\end{figure}

\clearpage

\begin{table}
\begin{center}
\caption{Measured Fluxes\label{tab1}\tablenotemark{a} }
\begin{tabular}{lllllll}
\tableline\tableline
Position\tablenotemark{b}& N$_{HI}$ & S$_{8}$ & S$_{24}$ & S$_{NUV}$ & S$_{H\alpha}$ & L(IR)\tablenotemark{c} \\
  &   (10$^{20}$cm$^{-2}$) & (mJy) & (mJy) & ($\mu$Jy) & 10$^{36}$~erg s$^{-1}$ & 10$^{40}$~erg s$^{-1}$ \\
\tableline
1 &  9.6 	 & 2.3$\pm$0.2     & 1.8$\pm$0.2      & 19$\pm$2       & (7.1$\pm$6.5) & 2.4 \\
2 &  4.7	 & 2.7$\pm$0.2     & 2.2$\pm$0.2      & 15$\pm$1       & (14$\pm$6)    & 2.9\\
3 &  6.8	 & 3.7$\pm$0.2     & 3.0$\pm$0.2      & 15$\pm$2       & 13$\pm$4      & 3.9  \\
4 &  3.6	 & 0.52$\pm$0.16   & 1.1$\pm$0.1      & 7.3$\pm$0.8    & 25$\pm$5      & 0.66\\
5 &  2.9	 & 1.3$\pm$0.1     & 1.3$\pm$0.1      & 12$\pm$1       & 22$\pm$5      & 1.4\\
6 &  4.3	 & 1.2$\pm$0.1     & 0.60$\pm$0.08    & 7.1$\pm$0.7    & 12$\pm$3      & 1.2  \\
7 &  4.0	 & 1.7$\pm$0.2     & 1.0$\pm$0.1      & 6.1$\pm$0.1    & 22$\pm$4      & 1.7  \\
8 &  5.1	 & 1.1$\pm$0.2     & 1.5$\pm$0.1      & 40$\pm$1       & (10$\pm$5)    & 1.4\\
9 &  5.0    & (0.49$\pm$0.17) & 0.65$\pm$0.16    & 4.6$\pm$1.1    & 18$\pm$2      & (0.58)\\
10 & 5.7    & 0.65$\pm$0.17   & 0.58$\pm$0.13    & (4.5$\pm$2.1)  & (4.4$\pm$4.6) & 0.71 \\
11 & 6.1    & 0.53$\pm$0.17   & (0.15$\pm$0.12)  & (1.3$\pm$2.1)  & 14$\pm$3      & (0.45)  \\ 
12 & 6.2    & 0.52$\pm$0.11   & (0.47$\pm$0.18)  & (7.6$\pm$4.8)  & (4.5$\pm$3.0) & (0.57)\\
13 & 5.0    & 0.48$\pm$0.12   & (0.44$\pm$0.19)  & (-0.2$\pm$13.2)& 24$\pm$3      & (0.53) \\
14 & 10     & 1.6$\pm$0.1     & 1.2$\pm$0.2      & 41$\pm$1       & 53$\pm$3      & 1.7\\
15 & 4.9    & 0.43$\pm$0.11   & 0.48$\pm$0.14    & (3.7$\pm$1.4)  & 20$\pm$4      & 0.49\\
16 & 5.8    & 0.40$\pm$0.06   & (0.09$\pm$0.15)  & 28$\pm$1       & 38$\pm$6      & (0.33)\\
17 & 5.0    & 0.38$\pm$0.10   & 0.30$\pm$0.09    & 17$\pm$2       & 16$\pm$3      & 0.40\\
18 & 3.7    & 0.26$\pm$0.06   & (0.78$\pm$0.28)  & 5.8$\pm$1.6    & 15$\pm$4      & (0.36)\\
19 & 5.9    & 0.52$\pm$0.08   & 1.0$\pm$0.2      & 13$\pm$2       & 17$\pm$4      & 0.66\\
20 & 4.5    & 1.0$\pm$0.3     & 0.83$\pm$0.19    & 23$\pm$2       & 18$\pm$3      & 1.1\\
21 & 5.6    & 3.7$\pm$0.2     & 1.4$\pm$0.2      & 41$\pm$2       & (12$\pm$5)    & 3.4\\

\tableline
\end{tabular}
\tablenotetext{a}{All values presented in units of flux per 21\arcsec\
  gaussian beam. Upper limits are enclosed in parentheses.}
\tablenotetext{b}{Positions in clockwise order, with offsets given in Table 2.}
\tablenotetext{c}{Values of L(IR) were derived from 8 and 24\micron\ fluxes
 using the calibration of Calzetti et al. (2005).}
\end{center}
\end{table}

\begin{deluxetable}{lllllllll}
\tabletypesize{\footnotesize}
\tablenum{2}
\tablecolumns{9}
\tablecaption{ Multiwavelength Star Formation Rates and Efficiencies\label{tab2}\tablenotemark{a}}
\tablehead{
\colhead {$\Delta\alpha$(\arcsec) \tablenotemark{b}} & \colhead{$\Delta\delta$(\arcsec)\tablenotemark{b}} & \colhead{M$_{gas}$\tablenotemark{c}}
& \colhead{SFR(H$\alpha$)} & \colhead{SFR(H$\alpha$,corr)\tablenotemark{d}} & \colhead{SFR(NUV)}
&  \colhead{SFR(L(IR))} & \colhead{SFE(H$\alpha$,corr)} & \colhead{SFE(L(IR))}\\
 \colhead{(\arcsec)}& \colhead{(\arcsec)} & \colhead{10$^{6}$M$_\odot$} & \colhead{10$^{-4}$M$_\odot$~yr$^{-1}$} &
\colhead{10$^{-4}$M$_\odot$~yr$^{-1}$} & \colhead{10$^{-4}$M$_\odot$~yr$^{-1}$} & \colhead{10$^{-4}$M$_\odot$~yr$^{-1}$} & \colhead{Gyr$^{-1}$} & \colhead{Gyr$^{-1}$}}
\startdata\\
\multicolumn{9}{c}{Outer Disk} \\
-40.5  &  225.8  &  11   & (0.56) & (0.80) & 3.1     & 11    &   (0.007)  & 0.097 \\
-63.8  &  209.3  &  5.5. & (1.1)  & (1.3)  & 2.5     & 13    &   (0.024)  & 0.23 \\
-90.8  &  162.0  &  8.0  & 1.0    & 1.3    & 2.5     & 18    &   0.016    & 0.22 \\ 
-207.8 &  113.3  &  4.2  & 2.0    & 2.3    & 1.2     & 3.0   &   0.053    & 0.070 \\
-189.8 &   65.3  &  3.4  & 1.7    & 1.9    & 2.0     & 6.5   &   0.056    & 0.19\\
-258.0 &  -52.5  &  5.1  & 0.95   & 1.1    & 1.2     & 5.2   &   0.022    & 0.10\\
-224.3 &  -198.8 &  4.7  & 1.7    & 2.0    & 1.0     & 7.6   &   0.043    & 0.16\\
-150.8 &  -254.3 &  6.0  & (0.79) & (0.95) & 6.5     & 6.3   &   (0.016)  & 0.10\\
-191.3 &  -270.8 &  5.9  & 1.4    & 1.7    & 0.75    & (2.6) &   0.029    & 0.044\\ 
-153.0 &  -310.5 &  6.7  & (0.35) & (0.43) & (0.73)  & 3.2   &   (0.006)  & (0.047)\\
-114.8 &  -342.8 &  7.2  & 1.1    & 1.4    & (0.21)  & (2.1) &   0.019    & (0.028)\\
 71.3  &  -333.8 &  7.3  & (0.36) & (0.45) & (1.2)   & (2.6) &   (0.006)  & (0.035)\\
 96.8  &  -306.8 &  5.9  & 1.9    & 2.3    & (0)     & (2.4) &   0.039    & (0.040)\\
 102.0 &  -273.8 &  12   & 4.2    & 6.0    & 6.7     & 7.6   &   0.051    & 0.064\\
 141.0 &  -267.8 &  5.8  & 1.6    & 1.9    & (0.60)  & 2.2   &   0.033    & 0.038\\
 221.3 &  -234.0 &  6.8  & 3.0    & 3.7    & 4.6     & (1.5) &   0.054    & (0.022)\\
 177.8 &  -225.8 &  5.9  & 1.3    & 1.5    & 2.8     & 1.8   &   0.026    & 0.031\\
 192.0 &  -150.8 &  4.4  & 1.2    & 1.4    & 0.95    & (1.6) &   0.031    & (0.038)\\
 201.0 &  -122.3 &  7.0  & 1.3    & 1.7    & 2.1     & 3.0   &   0.024    & 0.043\\
 208.5 &  -63.8  &  5.3  & 1.4    & 1.7    & 3.8     & 4.8   &   0.032    & 0.091\\
 204.8 &   90.0  &  6.6  & (0.95) & (1.2)  & 6.7     & 15    &   (0.018)  & 0.23\\
\multicolumn{9}{c}{Inner Disk} \\ 
-8.3    &  52.5  & 23.8  & 288  & 603 &  144  & 772 & 2.5  & 3.2\\ 
-63.8   &  37.5  & 24.5  & 126  & 269 &  138  & 499 & 1.1  & 2.0\\ 
-66.0   &  108.0 & 17.3  & 183  & 313 &  225  & 366 & 1.8  & 2.1\\ 
-102.0  & -15.8  & 11.6  & 90.1 & 129 &  227  & 321 & 1.1  & 2.8\\ 
-90.8   & -90.8  & 19.8  & 354  & 655 &  222  & 756 & 3.3  & 3.8\\ 
 38.3   & -64.5  & 17.8  & 132  & 229 &  161  & 672 & 1.3  & 3.8\\
 69.8   & -38.3  & 18.4  & 122  & 215 &  252  & 609 & 1.2  & 3.3\\ 
-159.8  & -128.3 & 19.3  & 68.5 & 125 &  58.5 & 163 & 0.64  & 0.84\\
 99.0   & -120.8 & 16.2  & 284  & 470 &  152  & 451 & 2.9  & 2.8\\ 
 142.5  & -14.3  & 18.4  & 107  & 189 &  139  & 222 & 1.0  & 1.2\\ 
\enddata
\tablenotetext{a}{Values enclosed in parentheses were derived from
  flux upper limits; see Table 1.}
\tablenotetext{b}{Offsets measured from the center position
  ($\alpha$,$\delta$)(J2000)=13:29:52.6,+47:11:43.8}
\tablenotetext{c}{Gas mass includes HI for the outer disk and (HI+H$_2$)
  for the inner disk. See \S\ref{sfrs} for details.}
\tablenotetext{d}{SFR derived from extinction-corrected H$\alpha$ fluxes.  See \S\ref{sfrs} for details.}

\end{deluxetable}


\begin{thebibliography}{}

\bibitem[Bianchi et al. (2005)]{bian05} Bianchi, L. et al. 2005, \apj,
  619, L71
\bibitem[Boulanger et al. 1998]{boul98} Boulanger, F., Boissel, P., Cesarsky, D., \& Ryter, C. 1998, \aap, 339, 194
\bibitem[Braine \& Herpin (2004)]{bh2004} Braine, J., \& Herpin, F. 2004, Nature, 432, 369
\bibitem[Bresolin et al. (2004)]{bre04} Bresolin, F., Garnett, D.R., \& Kennicutt, R.C. Jr. 2004, \apj, 615, 228
\bibitem[Calzetti et al. 2005]{calz05} Calzetti, D. et al. 2005, \apj, 633, 871
\bibitem[Dale et al. 2005]{dale05} Dale, D.A. et al. 2005, \apj, 633, 857
\bibitem[Engelbracht et al. 2005]{eng05} Engelbracht, C.W., Gordon,
  K. D., Rieke, G. H., Werner, M. W., Dale, D. A.; Latter, W. B. 2005,
  \apj, 628, L29
\bibitem[Fazio et al. (2004)]{faz04}Fazio, G. G., et al. 2004, \apjs,
  154, 10
\bibitem[Haynes et al. 1984]{hgc84}Haynes, M.P., Giovanelli, R., \& Chincarini, G.L. 1984,  \araa, 22, 445
\bibitem[Helfer et al. 2003]{helf03} Helfer, T.T., Thornley, M.D.,
  Regan, M.W., Wong, T., Sheth, K., Vogel, S.N., Blitz, L., \& Bock,
  D.C.-J. 2003, ApJS, 145, 259
\bibitem[Helou et al. (2000)]{hel00}Helou, G., Lu, H.Y., Werner, M.W., Malhotra, S., \& Silbermann, N.A. 2000, \apj, 532, L21
\bibitem[Helou et al. 2004)]{hel04} Helou, G., et al. 2004, ApJS, 154, 253
\bibitem[Kennicutt 1998]{kenn98} Kennicutt, R. C., Jr. 1998, \araa, 36, 189
\bibitem[Kennicutt et al. (2003)]{kenn03} Kennicutt, R. C., Jr., et al. 2003, PASP, 115, 928
\bibitem[Kewley et al. (2002)]{kew02} Kewley, L.J., Geller, M.J., Jansen, R.A., \& Dopita, M.A. 2002, \aj, 124, 3135
\bibitem[Leger \& Puget (1984)]{lp84} Leger, A. \& Puget, J.-L. 1984, A\&A, 137, L5
\bibitem[Lu et al. (2003)]{lu03} Lu, N., et al. 2003, \apj, 588, 199
\bibitem[Neff et al. (2005)]{neff05} Neff, S. G., et al. 2005, ApJL, 619, L91
\bibitem[Pahre et al. (2004a)]{pahre04a} Pahre, M.A., Ashby, M.L.N.,
  Fazio, G.G., \& Willner, S.P. 2004a, \apjs, 154, 229
\bibitem[Pahre et al. (2004b)]{pahre04b} Pahre, M.A., Ashby, M.L.N.,
  Fazio, G.G., \& Willner, S.P. 2004b, \apjs, 154, 235
\bibitem[Puget \& Leger (1989)]{pl89} Puget, J.-L. \& Leger, A. 1989, \araa, 27, 161
\bibitem[Rand et al. 1992]{rand92} Rand, R.J., Kulkarni, S.R., \&
  Rice, W. 1992, \apj, 390, 66
\bibitem[Regan et al. (2004)]{mwr04} Regan, M.W. et al. 2004, \apjs,
  154, 204
\bibitem[Rieke et al. 2004]{rieke04} Rieke, G.H. et al. 2004, \apjs,
  154, 25
\bibitem[Rots et al. (1990)]{rots90}Rots, A.H., Bosma, A., van der Hulst, J.
M., Athanassoula, E., \& Crane, P.C. 1990, AJ, 100, 387
\bibitem[Roussel et al. 2001]{rous01} Roussel, H., Sauvage, M.,
  Vigroux, L., Bosma, A. 2001, \aap, 372, 427
\bibitem[van de Hulst et al. 1954]{vdHMO54} van de Hulst, H.C.,
Muller, C.A., \& Oort, J.H. 1954, BAN, 12, 117
\bibitem[van der Kruit \& Allen 1978]{vdKA78} van der Kruit, P.C. \&
  Allen, R.J. 1978, \araa, 16, 103
\bibitem[Werner et al. 2004]{wern04} Werner, M.W. et al. 2004, \apjs,
  154, 1
\end{thebibliography}
\end{document}